\documentclass[twocolumn,prl,aps,showpacs]{revtex4-1}

\usepackage{graphicx}
\usepackage{color}
\usepackage[T1]{fontenc}
\usepackage{bm}
\usepackage{braket}

\usepackage{amsfonts}
\usepackage[utf8]{inputenc}
\usepackage{float}
\usepackage{amsmath}

\usepackage[normalem]{ulem}

\renewcommand{\vec}[1]{{\boldsymbol{#1}}} 

\usepackage{cancel}

\begin{document}
	
	
	\title{Finite-range effects in ultradilute quantum drops}
	\author{V. Cikojevi\'{c}, L. Vranje\v{s} 
		Marki\'{c}}
	\affiliation{Faculty of Science, University of Split, Ru\dj era 
		Bo\v{s}kovi\'{c}a 33, HR-21000 Split, Croatia}
	\author{J. Boronat}
	\affiliation{Departament de F\'{\i}sica, 
		Universitat Polit\`ecnica de Catalunya, 
		Campus Nord B4-B5, E-08034 Barcelona, Spain}
	\date{\today}
	
	\begin{abstract}
	In the first experimental realization of dilute Bose-Bose liquid drops 
using two hyperfine states of $^{39}$K some discrepancies between theory and 
experiment were observed. The standard analysis of the data 
using the Lee-Huang-Yang beyond mean-field theory predicted critical numbers 
which were significantly off the experimental measurements. Also, the radial 
size of the drops in the experiment proved to be larger than expected 
from this theory. Using a new functional, which is based on quantum Monte Carlo 
results of the bulk phase incorporating finite-range effects, we can explain 
the origin of the discrepancies in the critical number. This result proves the 
necessity of including finite-range corrections to deal with the observed 
properties in this setup. The controversy on the radial size is reasoned in 
terms of the departure from the optimal concentration ratio between the two 
species of the mixture.  
	\end{abstract}
	
	\maketitle
	
	\section{\label{intro} Introduction}

In the last years, there has been  an increasingly high interest 
in understanding dilute, ultracold quantum Bose-Bose mixtures. 
The focus on this study increased dramatically after the theoretical 
proposal by Petrov~\cite{petrov} on the formation of self-bound liquid drops.
These liquid drops are stabilized by beyond-mean-field effects and can appear 
in mixtures of two Bose-Einstein condensates with 
repulsive intraspecies and attractive interspecies interactions.
The drops originate from a delicate balance between the collapsed state, 
predicted by 
mean-field (MF) theory, and the repulsive character of the 
first beyond mean-field term (Lee-Huang-Yang 
-LHY-). The same perturbative theoretical scheme predicts self-binding in 
low-dimensional mixtures~\cite{petrov_astra2D, parisi} and dipolar 
systems~\cite{dipolar_experiment_nature, raul}. 
Recently, these predicted quantum drops have been observed in several 
experiments~\cite{bb_mixture_first_taruell,soliton_to_drop,bb_mixture_sec, bb_mixture_het} and they resemble the well-known liquid Helium drops~\cite{barranco_review, 
krotscheck_drops}. However, the inner 
density in the Bose-Bose drops is about five orders of magnitude  
smaller than in $^4$He~\cite{barranco_review,krotscheck_drops}.
Therefore, these new quantum drops extend the realm of the liquid state to 
much lower densities than any previous existing classical or quantum  
liquid.

In the two labs~\cite{bb_mixture_first_taruell,bb_mixture_sec} where the drops 
have been observed, the Bose-Bose mixture is composed of two hyperfine states 
of 
$^{39}$K. In the first experiment by  Cabrera \textit{et 
al.}~\cite{bb_mixture_first_taruell}, the drops are harmonically confined in 
one of the directions of space whereas in the second one by Semeghini 
\textit{et al.}~\cite{bb_mixture_sec} the drops are observed in free space. 
This difference in the setup makes that in the first case the drops are not 
spherical like in the second experiment. This also affects the critical 
number, that is, the minimum number of atoms required to get a self-bound 
state. The measured critical numbers differ significantly between the two labs 
due to the different shape of the drops, the ones in the confined case being 
smaller than in the free case. In both works, the experimental results for the 
critical number are compared with the MF+LHY theory. The agreement between this 
theory and the drops produced in free space is quite satisfactory in spite of 
the large errorbars of the experimental data that hinder a precise comparison. 
However, in the confined drops of Ref.~\cite{bb_mixture_first_taruell}, where 
the critical numbers are significantly smaller than in the free case,  the 
theoretical predictions do not match well the experimental data.

Ultradilute liquid drops, which require beyond-mean field corrections to 
be theoretically understood, offer the perfect benchmark to explore possible 
effects beyond MF+LHY theory~\cite{jorgensen} 
which usually play a minute role in the case of single-component 
gases~\cite{pethick,giorgini_boronat}. Indeed, several theoretical  
studies~\cite{staudinger, symmetric, tononi1, tononi2,salsnich_nonuniversal} 
indicate a strong dependence of the equation of state of the liquid on the 
details of the interatomic interaction, even at very low densities. This essentially means it is already possible to achieve observations outside the universal regime, in which all the interactions can be expressed in terms of the gas parameter $n a^3$, with $a$ the s-wave scattering length. The first correction 
beyond this universality limit must incorporate the next term in the scattering series, that is the effective range $r_{\rm eff}$~\cite{scattering_theory_of_waves_and_particles_1, scattering_theory_of_waves_and_particles_2}, which in fact can be 
quite  large in these drops and in alkali atoms 
in general~\cite{flambaum,effective_range_taruell}. 


	Motivated by experiments with quantum drops, we have investigated 
the self-bound quantum mixture composed of two hyperfine states of $^{39}$K 
using  nonperturbative quantum Monte Carlo (QMC) methods. Direct QMC 
simulations~\cite{cikojevic} of finite particle-number drops, as produced in 
experiments, would serve as a great test of mean-field theory but, 
unfortunately, this is not yet achievable because of the large number of 
particles in realistic drops ($N>10^4$).  Yet, 
the problem can be addressed in the Density Functional Theory (DFT) spirit, 
relying on the Hohenberg-Kohm-Sham 2nd theorem~\cite{kh_theorem}, which 
guarantees that a density functional exists that matches exactly the 
ground-state solution.
To build a functional for the quantum Bose-Bose mixture, we have 
carried out calculations in bulk conditions using the diffusion Monte Carlo 
(DMC) method, an exact QMC technique applicable to systems at zero temperature. 
Using that functional we can access to energetics and structure of 
the liquid drops in the same conditions as in the experiment. We focus on the 
data obtained by Cabrera \textit{et al.}\cite{bb_mixture_first_taruell} in the 
confined setup since it is in that case where discrepancies between MF+LHY 
theory were observed. 

The rest of the paper is organized as follows. In Sec. II, we introduce the 
theoretical methods used for the study and discuss the way in which the density 
functional is built. Sec. III comprises the results obtained for the bulk 
liquid using the available scattering data of the $^{39}$K mixture. The 
inclusion of the effective range parameters in the interaction model allows for 
a better agreement with the measured critical numbers. Finally, we summarize 
the most relevant results here obtained and derive the main conclusions of our 
work.

	\section{\label{methods} Methods}
		
	We study a mixture of two hyperfine states of $^{39}$K bosons at zero 
temperature. The Hamiltonian of the system is
	\begin{equation}
		H = \sum_{i=1}^{N}-\dfrac{\hbar^2}{2m}\nabla_i^2 + \dfrac{1}{2}   
\sum_{\alpha, \beta=1}^{2} \sum_{i_\alpha, j_\beta=1}^{N_\alpha 
N_\beta}V^{(\alpha, \beta)}(r_{i_\alpha j_\beta}) \ ,
\label{hamiltonian}
\end{equation}
where $V^{(\alpha, \beta)}(r_{i_\alpha j_\beta})$ is the interatomic 
potential between species $\alpha$ and $\beta$. The mixture is composed of 
$N=N_1 + N_2$ atoms, with $N_1$ ($N_2$) bosons of type 1 (2). The potentials 
are chosen to reproduce the experimental scattering parameters, and 
we have used different model potentials to investigate the influence of the 
inclusion of the effective 
range. The microscopic study has been carried out using  
a second-order DMC method~\cite{dmc}, which allows for an exact 
estimation of the ground-state of the mixture within some statistical 
errors. DMC solves stochastically the imaginary-time Schr\"odinger equation 
using a trial wave function as importance sampling which guides the diffusion 
process to regions of expected large probability. In the present case, we 
used a trial wavefunction built as a product of Jastrow factors 
\cite{jastrow},
	\begin{equation}
		\Psi(\vec{\mathrm{R}}) = \prod_{i<j}^{N_1} f^{(1,1)}(r_{ij}) 
\prod_{i<j}^{N_2} f^{(2,2)}(r_{ij}) \prod_{i,j}^{N_1, N_2} f^{(1,2)}(r_{ij}) \ ,
	\end{equation} 
where the two-particle correlation functions $f(r)$ are 
	\begin{equation}
	f^{\alpha, \beta}(r)=
	\begin{cases}
	f_{\rm 2b}(r) &  r < R_0 \\
	B\exp(-\frac{C}{r} + \frac{D}{r^2}) 	  ,& R_0 <r < L/2 \\
	1	  ,& r > L/2 \ . \\
	\end{cases}
	\end{equation}
	The function
$f_{\rm 2b}$ is the solution of the two-body problem for a specific 
interaction model, $R_0$ is a variational parameter, and $L=(N / 
\rho)^{1/3}$ is the size of the simulation box. There is a weak dependence of 
the variational energy on $R_0$, and it has been kept as $R_0 = 0.9 L/2$ for 
all the cases.
A careful analysis of imaginary time-step dependence and population 
size bias has been carried out, keeping both well under the statistical error. 
The time-step dependence is well eliminated for $\Delta \tau = 0.2 \times 
ma_{11}^2 / \hbar$ and the population bias by using $n_{\rm w} = 100$. 
Our simulations  are performed in a 
cubic box with periodic boundary condition, using a number of particles $N$. 
The thermodynamic limit is achieved by repeating calculations with different 
particle numbers; we observe that, within our numerical precision, the energy 
per particle converges at $N\approx 600$ for the range of magnetic 
fields here considered.

Within density functional theory (DFT), we seek for a 
many-body wave function built as a product of single-particle orbitals, 
	\begin{equation}
		\label{eq:dft_wf}
		\Psi(\vec{r}_1, \vec{r}_2, \ldots, \vec{r}_N)= 
		\prod_{i=1}^{N} \psi(\vec{r}_i).
	\end{equation}	
	These single-particle wave functions, which in general are time-dependent, 
are obtained by solving the Schr\"odinger-like equation~\cite{francesco_manuel},
	\begin{equation}
		\label{eq:time_dep_gp}
		i\hbar\dfrac{\partial \psi}{\partial t} = \left(-\dfrac{\hbar^2}{2m} 
\nabla^2 + V_{\rm ext}(\vec{r}) + \dfrac{\partial \mathcal{E}_{\rm 
int}}{\partial \rho}\right) \psi \ ,
	\end{equation}
	where $V_{\rm ext}$ is an external potential acting on the system and
$\mathcal{E}_{\rm int}$ is an energy per volume term that accounts for the 
interparticle correlations.	 The differential equation 
(\ref{eq:time_dep_gp}) is solved by propagating the wave function $\psi$ 
with the time-evolution operator
	\begin{equation}
		\psi(t + \Delta t) = e^{-i H \Delta t} \psi(t) \ .
	\end{equation}
	To this end, we have implemented a three-dimensional numerical solver based 
on the Trotter decomposition of the time evolution 
operator~\cite{chin_krotchek,manuel_marti} with second-order accuracy  
in the timestep $\Delta t$, as follows 
	\begin{equation}
	e^{-i H \Delta t}  = e^{-i\Delta t V(\vec{R}') / 2}  e^{-i\Delta 	t 	K } e^{-i\Delta t V(\vec{R})/ 2} + \mathcal{O}(\Delta t^2) \ ,
	\end{equation}
	with $K$ and $V$ the kinetic and potential terms in Eq. 
\ref{eq:time_dep_gp}.

	\section{\label{results} Results}
	
\begin{figure}[tb]
		\centering
		\includegraphics[width=\linewidth]{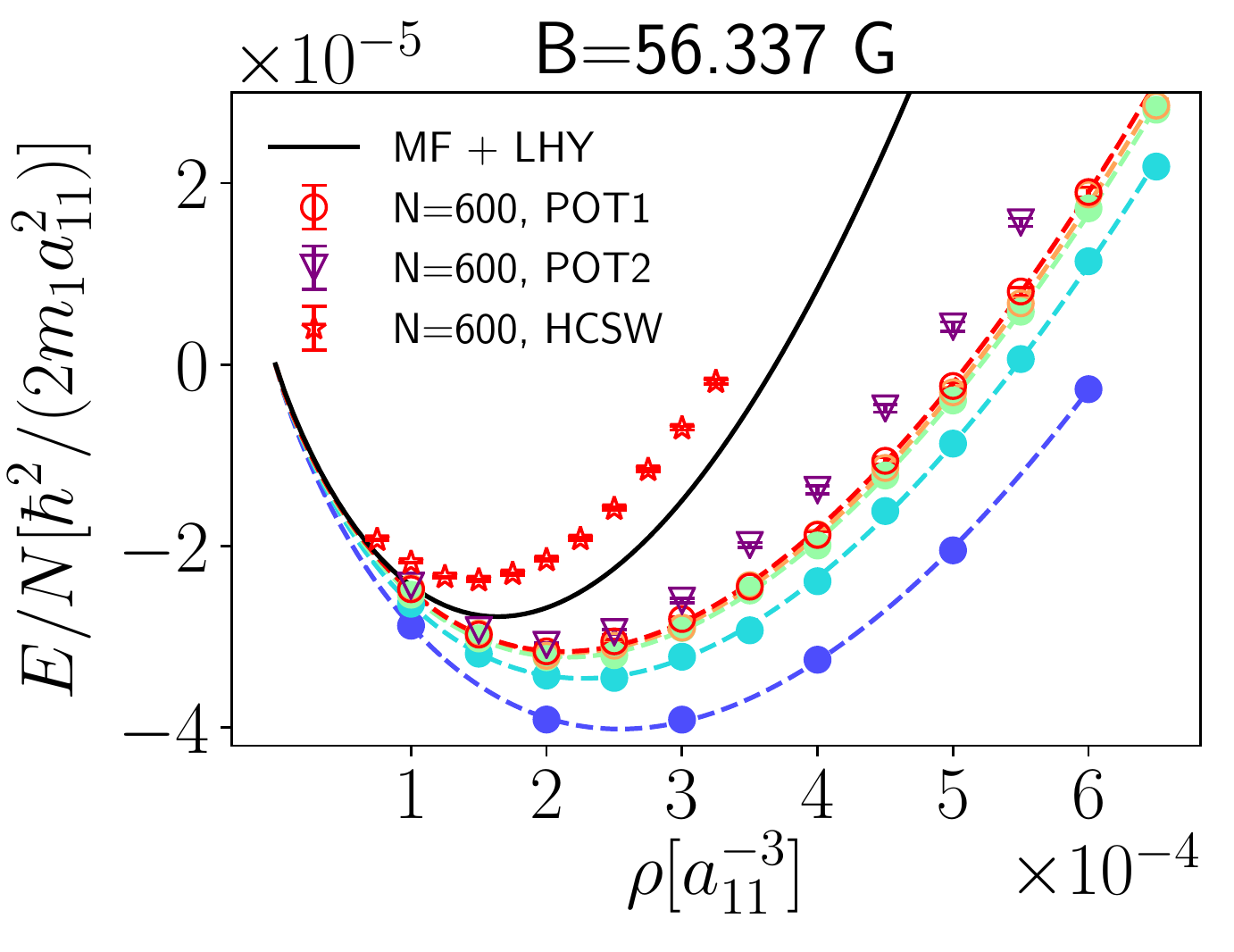}
		\caption{Dependence of the equation of state on the effective range for selected potential models, compared with MF+LHY. Full circles are calculations using POT1, and we illustrate the convergence to negligible finite-size effects starting from $N=100$ (lower points), 
200, 400, 500 to $N=600$ (upper points). Dashed lines are fits to the DMC data  
with Eq.(\ref{eq:density_functional}).}
		\label{fig:ploteosfinitesizeb56337}
	\end{figure}

	In order to go beyond the MF+LHY density functional we have carried out DMC 
calculations of the bulk liquid. In the 
mixture of $^{39}$K under study, we call the state $\ket{F,m_{\rm 
F}}=\ket{\downarrow} = 
\ket{1,0}$ as component 1, and the state $\ket{F,m_{\rm F}}=\ket{\uparrow} = 
\ket{1,-1}$ as component 2.
In Fig.~\ref{fig:ploteosfinitesizeb56337}, 
we show the energy per particle of the $^{39}$K mixture as a function of the 
density, 
using three different sets of potentials in the Hamiltonian 
(\ref{hamiltonian}): 
\begin{itemize}

\item[i)] Hard-core interactions (HCSW) with diameter 
$a_{ii}$, $i=1,2$, for the repulsive intraspecies interaction, and a 
square-well potential 
with range $R=a_{11}$ and depth $V_0$ for the interspecies 
potential. The three potentials reproduce the s-wave scattering lengths for the 
three channels,

\item[ii)] POT1 stands for a set of potentials which reproduces both the 
s-wave scattering lengths and effective ranges of the three interacting 
pairs of the $^{39}$K mixture.  To model the interactions, we have chosen a 
square-well 
square barrier potential~\cite{sssb_potential} for the $11$ channel, a 
10-6 Lennard-Jones potential~\cite{pade} for the $22$ channel, and a 
square-well 
potential of range $R$ and depth $V_0$~\cite{pethick} in the $12$ channel, 

\item[iii)] POT2 also reproduce both the s-wave scattering lengths and 
effective ranges, by using a sum of Gaussians in the $11$ channel, a 
10-6 Lennard-Jones potential in the $12$ channel, and finally a soft-sphere 
square well in the $22$ channel. 
\end{itemize}

In all cases, the attractive interatomic potential does not support a two-body 
bound state. We have obtained  the s-wave scattering length and effective range 
of the potentials using standard scattering 
theory~\cite{scattering_theory_of_waves_and_particles_1, scattering_theory_of_waves_and_particles_2}. 

We compare our DMC results to the 
MF+LHY theory, which can be compactly written as \cite{petrov}
	\begin{equation}
		\label{eq:mflhy_eos}
		\dfrac{E/N}{|E_0 / N|} = -3\left(\dfrac{\rho}{\rho_0}\right) + 2 
\left(\dfrac{\rho}{\rho_0}\right)^{3/2} \ ,
	\end{equation}
assuming the optimal concentration of particles from  
mean-field theory, $N_1 / N_2 = \sqrt{a_{22} / a_{11}}$. The energy per 
particle $E_0 / N$ at the equilibrium density of the MF+LHY approximation $\rho_0$  and $\rho_0$ itself is

	\begin{equation}
\label{eq:en_0}
E_0 / N = \dfrac{25 \pi^2 \hbar^2 |a_{12} + \sqrt{a_{11} 
		a_{22}}|^3}{768m a_{22} a_{11} \left(\sqrt{a_{11}} + \sqrt{a_{22}}\right)^6},
\end{equation} 
\begin{equation}
\label{eq:rho_0}
\rho_0 a_{11}^3 = \dfrac{25 \pi}{1024} \dfrac{\left(a_{12}/a_{11} + 
	\sqrt{a_{22}/a_{11}}\right)^2}{\left(a_{22}/a_{11}\right)^{3/2}\left(1+\sqrt{a_{
			22}/a_{11}}\right)^4} .
\end{equation}

\begin{table}[]
	\caption{Scattering parameters \cite{sclens}, s-wave scattering length $a$ and the effective range $r^{\rm eff}$ in units of Bohr radius $a_0$, as a function of the magnetic field $B$.  } 
	\begin{tabular}{ c | c | c | c | c | c | c }
		\hline
		$B (G)$ & $a_{11} (a_0)$ & $r_{11}^{\rm eff}(a_0)$ & $a_{22} (a_0)$ & $r_{22}^{\rm eff}(a_0)$ & $a_{12} (a_0)$ & $r_{12}^{\rm eff}(a_0)$ \\
		\hline 
		56.230 & 63.648           & -1158.872                                             & 34.587           & 578.412                                               & -53.435          & 1021.186                                              \\
		56.337 & 66.619           & -1155.270                                             & 34.369           & 588.087                                               & -53.386          & 1022.638                                              \\
		56.395 & 68.307           & -1153.223                                             & 34.252           & 593.275                                               & -53.360          & 1022.617                                              \\
		56.400 & 68.453           & -1153.046                                             & 34.242           & 593.722                                               & -53.358          & 1022.616                                              \\
		56.453 & 70.119           & -1150.858                                             & 34.136           & 599.143                                               & -53.333          & 1023.351                                              \\
		56.511 & 71.972           & -1148.436                                             & 34.020           & 604.953                                               & -53.307          & 1024.121                                              \\
		56.574 & 74.118           & -1145.681                                             & 33.895           & 610.693                                               & -53.278          & 1024.800                                              \\
		56.639 & 76.448           & -1142.642                                             & 33.767           & 616.806                                               & -53.247          & 1025.593    \\
	\end{tabular}
\end{table}

In Fig.~\ref{fig:ploteosfinitesizeb56337}, we report DMC results for the 
equation of state corresponding to a magnetic field $B=56.337$ G, one of the magnetic fields used in experiments. We show the convergence of 
the results on the number of particles in the simulation for the particular 
case of POT1 set of potentials. As we can see, the convergence is achieved 
with $N=600$. We have repeated this analysis for all the potentials and, in all 
the magnetic field range explored, we arrive to convergence with similar 
$N$ values. We have investigated the dependence on the 
effective range by repeating the calculation using the HCSW and POT2 
potentials. As 
it is clear from  Fig.~\ref{fig:ploteosfinitesizeb56337}, only when both 
scattering parameters, the s-wave scattering 
length and the effective range, are imposed on the model potentials 
we get an approximate universal equation of state, mainly around the 
equilibrium density. The equation of state so 
obtained shows a significant and overall decrease of the energy compared to the 
MF+LHY prediction, with a correction that increases with the density.
Instead, using the HCSW potentials, 
which only fulfill the s-wave scattering lengths, the energies obtained  
are even above the  MF+LHY prediction. A similar  
behavior has been previously shown to hold in symmetric ($N_1=N_2$) Bose-Bose 
mixtures~\cite{symmetric}.

	\begin{figure}[tb]
		\centering
		\includegraphics[width=\linewidth]{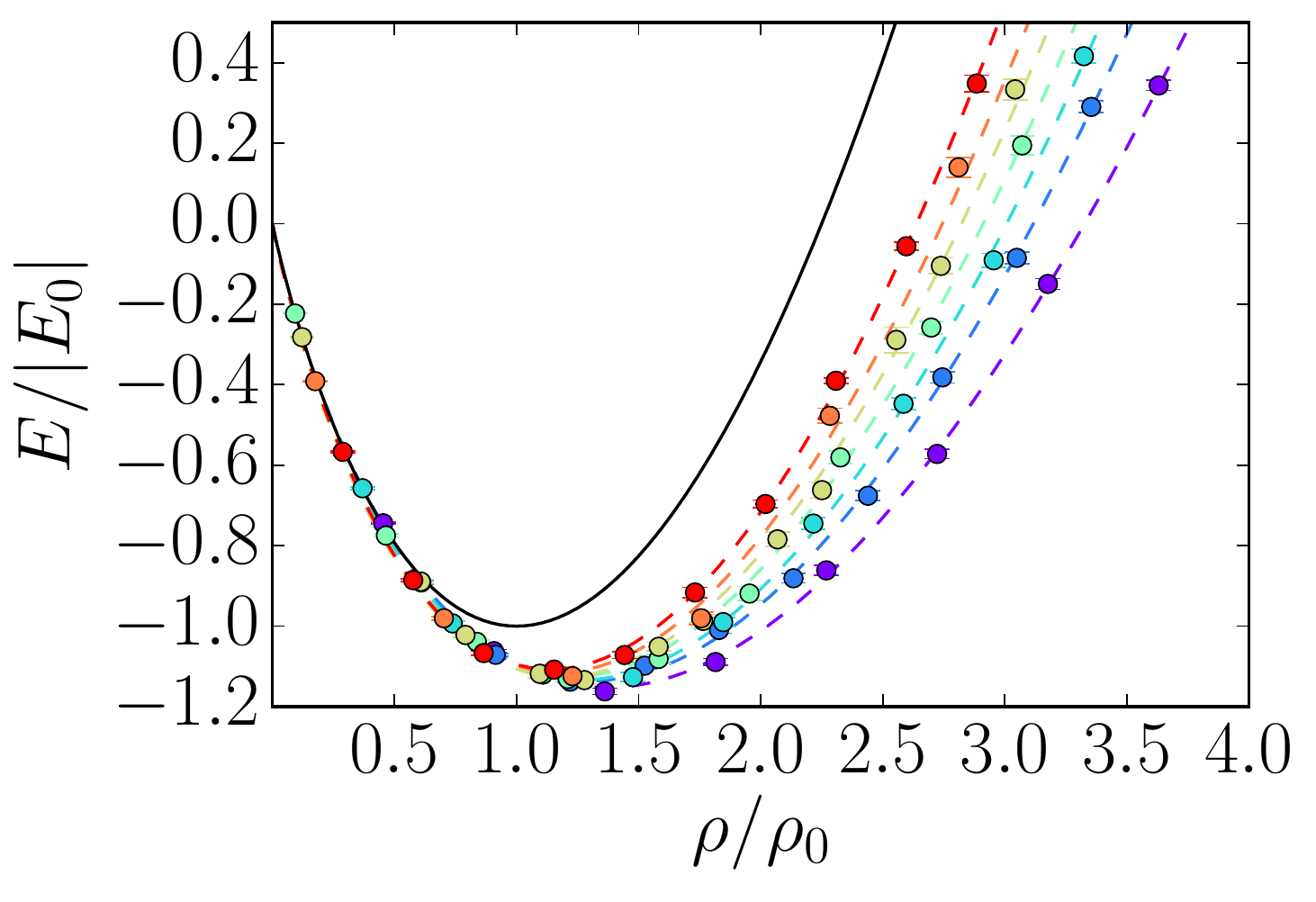}
		\caption{DMC energy per particle as a function of the density (circles), starting from $B=56.230 G$ (lower points) to $B=56.639 G$ (upper points). Energy and density are normalized to $E_0$ and $\rho_0$, given in Eq. \ref{eq:en_0} and  \ref{eq:rho_0}, respectively. Dashed lines are fits with Eq. (\ref{eq:density_functional}). Full line is the MF+LHY theory (Eq. \ref{eq:mflhy_eos}).}
		\label{fig:eoscombineddiffbuniversal}
	\end{figure}	

	Equations of state of the bulk mixture, for the seven values of the 
magnetic field used in the experiments ($B=56.230$ G  to $B=56.639$ G),  are 
shown in Fig.~\ref{fig:eoscombineddiffbuniversal}. The DMC results are 
calculated using the model POT1, but the differences with the other set POT2 
are not significant. In all cases, we take the mean-field prediction for the 
optimal ratio  of partial densities $\rho_1 / \rho_2 = \sqrt{a_{22} / a_{11}}$. 
We have verified in several cases that this is also the concentration 
corresponding to the ground state of the system in our DMC calculations, i.e., 
the one that gives the minimum energy at equilibrium.   
The DMC results are compared with the MF+LHY equation of 
state (\ref{eq:mflhy_eos}). Overall, a reduction of the 
magnetic field, or equivalently an increase in $|\delta a|=a_{12} + 
\sqrt{a_{11}a_{12}}$, leads to an increase of the binding energy
compared to  the MF+LHY approximation. This happens clearly due to the influence of the large experimental effective range, since in the limit of zero range
one would observe overall repulsive beyond-LHY terms (see also Fig. 
\ref{fig:ploteosfinitesizeb56337} and Ref.~\cite{symmetric}).	
	\begin{figure}[tb]
		\centering
 		\includegraphics[width=\linewidth]{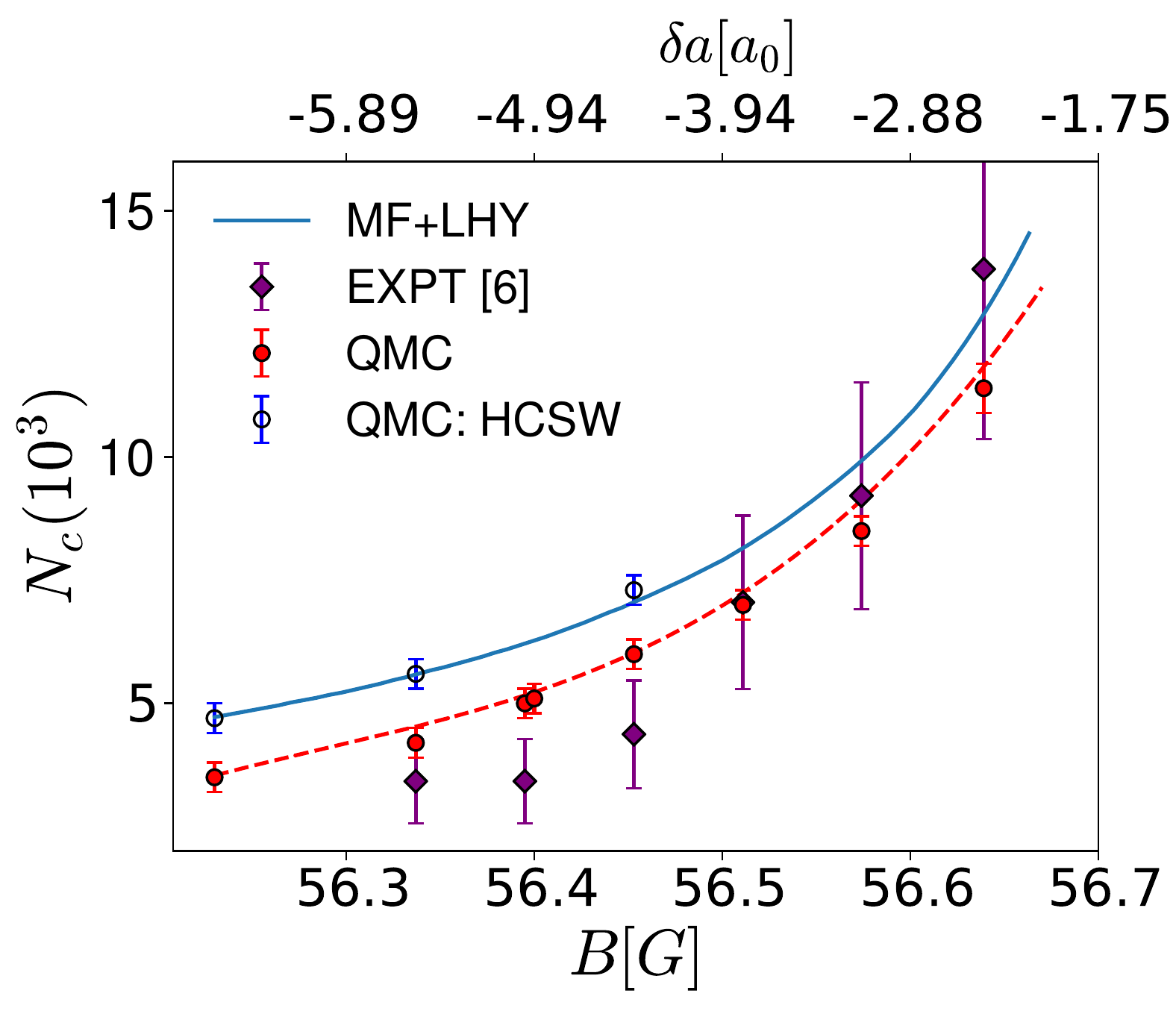}
		\caption{Dependence of the critical atom number on the 
		magnetic field. Full circles are predictions  using the QMC functional within DFT with the interaction potentials which reproduce both $a$ and $r_{\rm eff}$. Diamond points are data from the experiment \cite{bb_mixture_first_taruell}. Empty points show the prediction using the QMC functional with the HCSW model potentials. }
		\label{fig:criticalncomparisonwicfo}
	\end{figure}

\begin{table}[tb]
	\caption{Critical atom number to form a droplet in a harmonic trap $V_z = 
\frac{1}{2}m\omega_z^2 z^2$, where $a_{\rm ho} = \sqrt{\hbar / (m\omega_z)} = 
0.639 \, \mu m$ is the same value as in the 
experiment~\cite{bb_mixture_first_taruell}.  $\varepsilon_r= |N_c^{\rm 
QMC} - N_c^{\rm MFLHY}|/{N_c^{\rm MFLHY}}$ is the relative error.}
	\label{tab:nc_icfo} 
	\begin{tabular}{ c | c | c | c | c}
		\hline
		$B (G)$ & $N_c^{\rm QMC}$ & $N_c^{\rm MFLHY}$ &  
 $\varepsilon_r$ & $N_c^{\rm QMC} - 
N_c^{\rm MFLHY}$  \\
		\hline 
		56.23  & 3500  & 4650   & 0.25 & -1150 \\
		56.337 & 4200  & 5570   & 0.25 & -1370 \\
		56.395 & 5000  & 6200   & 0.19 & -1200 \\
		56.4   & 5100  & 6250   & 0.18 & -1150 \\
		56.453 & 6000  & 7000   & 0.14 & -1000 \\
		56.511 & 7000  & 8050   & 0.13 & -1050 \\
		56.574 & 8500  & 9800   & 0.13 & -1300 \\
		56.639 & 11300 & 12700  & 0.11 & -1400    \\
	\end{tabular}
\end{table}

\begin{table}[tb]
	\caption{Critical atom number for spherical free
drops~\cite{bb_mixture_sec}. $\varepsilon_r= |N_c^{\rm 
QMC} - N_c^{\rm MFLHY}|/{N_c^{\rm MFLHY}}$ is the relative error.
}
	\label{tab:nc_florence}
	\begin{tabular}{ c | c | c | c | c }
		\hline
		$B (G)$ & $N_c^{\rm QMC}$ & $N_c^{\rm MFLHY}$ & $\varepsilon_r$ & 
$N_c^{\rm QMC} - N_c^{\rm MFLHY}$  \\
		\hline 
		56.23  & 16000  & 15800  & 0.01  & 200   \\
		56.337 & 24600  & 24900  & 0.01 & -300  \\
		56.395 & 32700  & 33900  & 0.04 & -1200 \\
		56.4   & 35300  & 35500  & 0.01 & -200  \\
		56.453 & 47200  & 47700  & 0.01 & -500  \\
		56.511 & 69100  & 70600  & 0.02 & -1500 \\
		56.574 & 114000 & 119000 & 0.04 & -5000 \\
		56.639 & 230000 & 236000 & 0.03 & -6000    \\
	\end{tabular}
\end{table}

DMC energies for the $^{39}$K mixture are well fitted using the functional form 
	\begin{equation}
	\label{eq:density_functional}
	E/N = \alpha \rho  + \beta \rho^\gamma,
	\end{equation}
	as it can be seen in Fig.~\ref{fig:eoscombineddiffbuniversal}. These 
equations of state, calculated within the range of magnetic fields used in 
experiments, are then used in the functional form (\ref{eq:time_dep_gp}) with 
$\mathcal{E}_{\rm{int}}=\rho\,E/N$. With the new functional, based on our DMC 
results, we 
can study the quantum drops with the proper number of particles which is too 
large for a direct DMC simulation.

Results for the critical atom number $N_c$ at different $B$ are shown  in 
Fig.~\ref{fig:criticalncomparisonwicfo} in comparison with the  
experimental 
results of Ref.~\cite{bb_mixture_first_taruell}. To make the comparison 
reliable, we have included the same transversal confinement as in the 
experiment. In particular, theoretical predictions are obtained within DFT, using a Gaussian ansatz $\phi=\exp\left(-r^2 / (2\sigma_r^2) -z^2 / (2\sigma_z^2)\right)$. When the equation of state of the bulk takes into account the 
effective range of all the pairs we observe an overall decrease of $N_c$ with 
respect to the MF+LHY prediction. Interestingly, if we use  the HCSW model 
potentials, with essentially zero range, our results are on top of the  
MF+LHY line (see the points at $B=56.23$ G, $B=56.337$ G and $B=56.453$ G in 
Fig.~\ref{fig:criticalncomparisonwicfo}). The observed decrease of $N_c$ leads 
our theoretical prediction closer to the experimental data in a significant 
amount and in all the $\delta a$ range, clearly showing the significant 
influence of the effective range on the $N_c$ values. Experiments on quantum 
droplets were performed either in the harmonic 
trap~\cite{bb_mixture_first_taruell} or in a free-drop 
setup~\cite{bb_mixture_sec}. Predictions of $N_c$ for these two geometries are 
given in Tables~\ref{tab:nc_icfo} and \ref{tab:nc_florence}, using MF+LHY and 
QMC functionals. The absolute difference of predicted $N_c$ values between the 
two functionals are about 1000 atoms. On the other hand, the relative 
difference is much higher in the harmonically-trapped system because the 
presence of an external trap significantly reduces $N_c$. 
	\begin{figure}[tb]
		\centering
		\includegraphics[width=1.\linewidth]{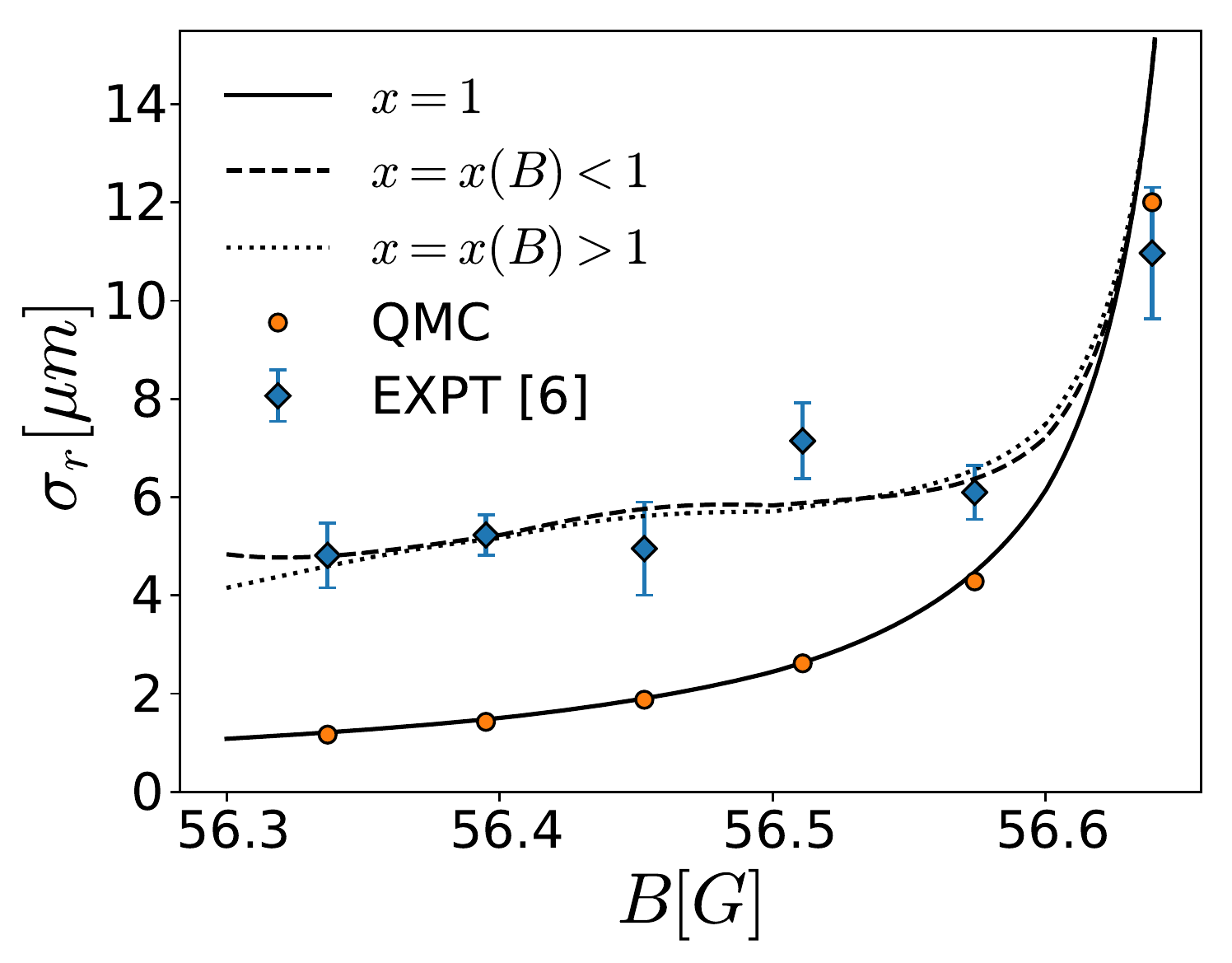}
		\caption{Dependence of the radial size of a $N=15000$ drop on the 
external magnetic field, or equivalently the residual s-wave scattering length. 
Lines are predictions under MF+LHY theory; full line is a prediction with 
$x=1$, 
 dashed and dotted lines are fits of experimental sizes using a parameter $x$.}
		\label{fig:sigmarcompwicfoxofb}
	\end{figure}

\begin{figure}[tb]
		\centering
		\includegraphics[width=1.\linewidth]{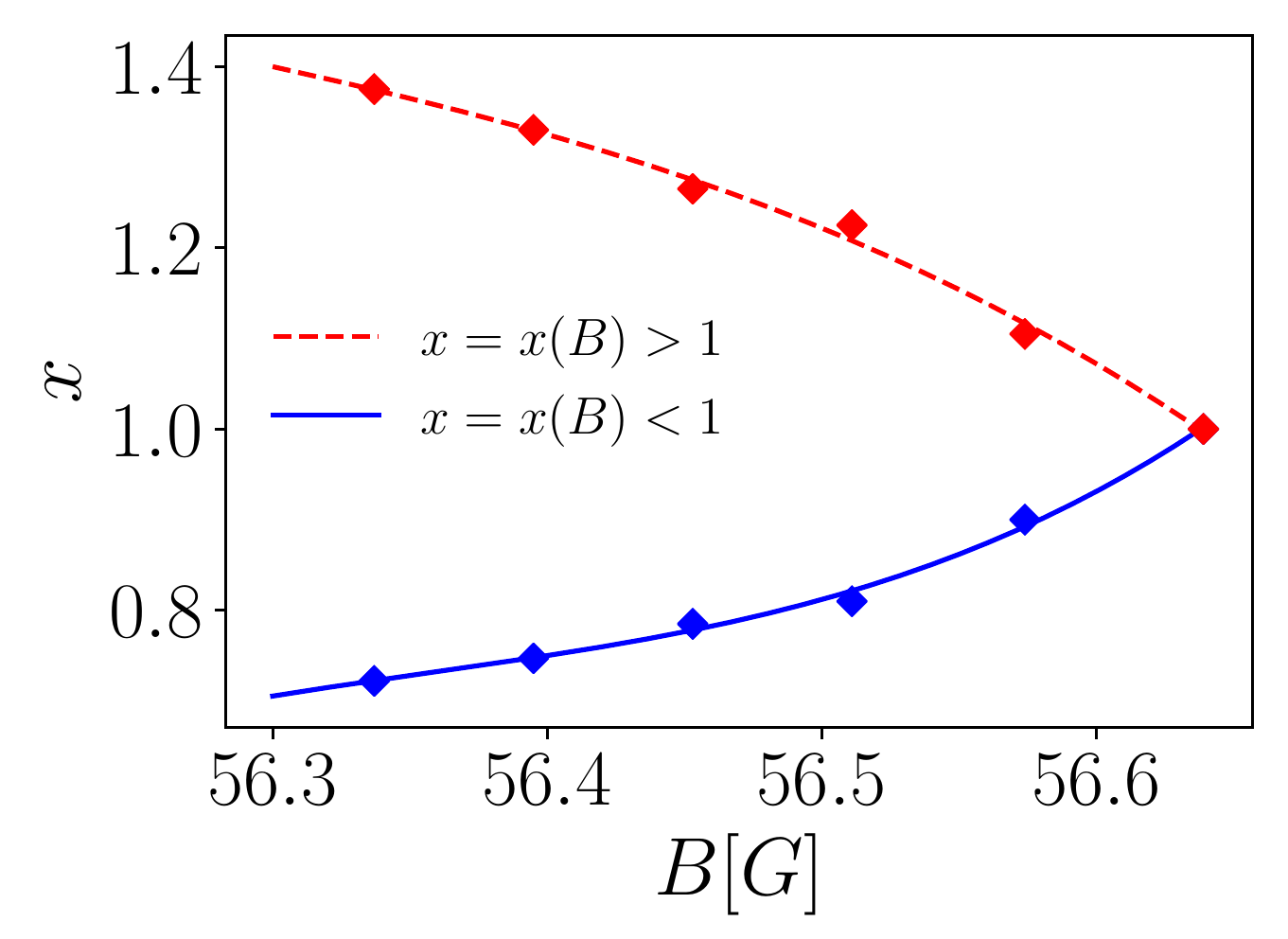}
		\caption{Values of $x=N_2 / N_1 \sqrt{a_{22} / a_{11}}$ which reproduce 
the experimental size of a N=15000 drop (Fig. \ref{fig:sigmarcompwicfoxofb}) 
within the MF+LHY theory, as a function of the magnetic field. Points are the 
values which reproduce the size, and lines are power-law fits of $x$ as a 
function of the magnetic field $B$. Note that two solutions exist since the 
choice of naming each component is twofold.}
		\label{fig:xofbfitcombined}
	\end{figure}

	\begin{figure}[tb]
		\centering
		\includegraphics[width=1.\linewidth]{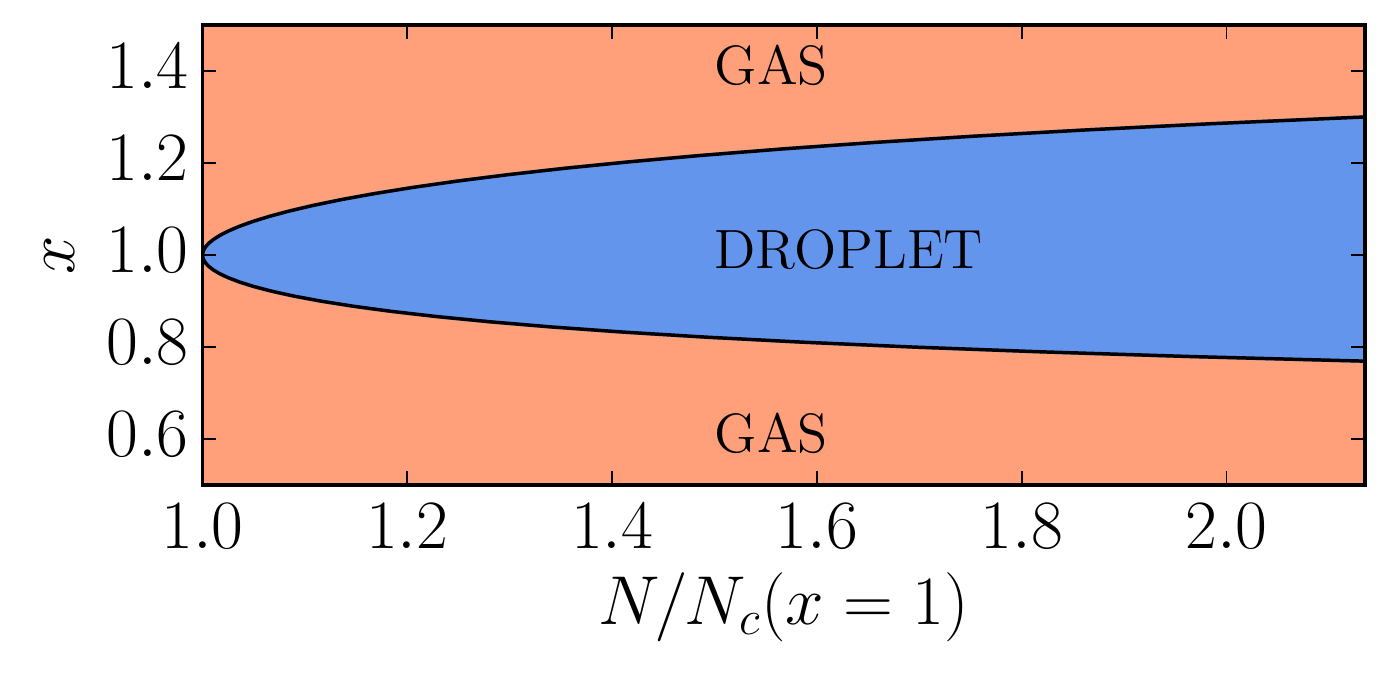}
		\caption{Phase diagram of $^{39}$K at $B=56.230$G using MF+LHY theory, 
spanned with $x = N_2 / N_1 \sqrt{a_{22} / a_{11}}$ and the total particle number 
$N$, normalized with the critical atom number $N_c$ evaluated at $x=1$ 
\cite{petrov}.}
		\label{fig:xnc}
	\end{figure}

A second observable measured in experiments is the size of the drops. 
The radial size of a $N=15000$ drop for different values of the magnetic 
field was reported in Ref.~\cite{bb_mixture_first_taruell}. In 
Fig.~\ref{fig:sigmarcompwicfoxofb}, we compare the experimental values with 
different theoretical predictions. We observe a slight reduction 
in size using QMC functionals, compared to MF+LHY theory, which is a 
consequence 
of the stronger binding produced by inclusion of finite range interactions. 
Since the 
experimental data go to the opposite direction, it means that drops size 
can not be 
explained solely in terms of the non-zero effective range. One 
possible explanation for this clear disagreement 
could be a deviation from the optimal relative number of particles, 
which can occur in non-equilibrated drops or  when one of the 
components has a large three-body recombination coefficient. Let us define 
$x=N_2 / N_1 \sqrt{a_{22} / a_{11}}$. Then, $x=1$ stands for the optimal 
relative particle number, i.e., the concentration corresponding to the 
ground-state of the system. We have investigated the behavior of both the 
MF+LHY 
and QMC functionals under variations in $x$, and both predict a decrease in 
the drop size proportional to the deviation from $x=1$. Using the MF+LHY 
functional, we have obtained the $x$ values that fit the experimental size for 
every $B$ (Fig.~\ref{fig:sigmarcompwicfoxofb}). We report the result of this analysis in Fig.~\ref{fig:xofbfitcombined}; notice that there is a symmetry on $x$ and so only its absolute deviation from one is important. This result clearly shows the sensitive dependence of drop structural properties on the relative atom number.

	\begin{figure}[tb]
		\centering
		\includegraphics[width=1.\linewidth]{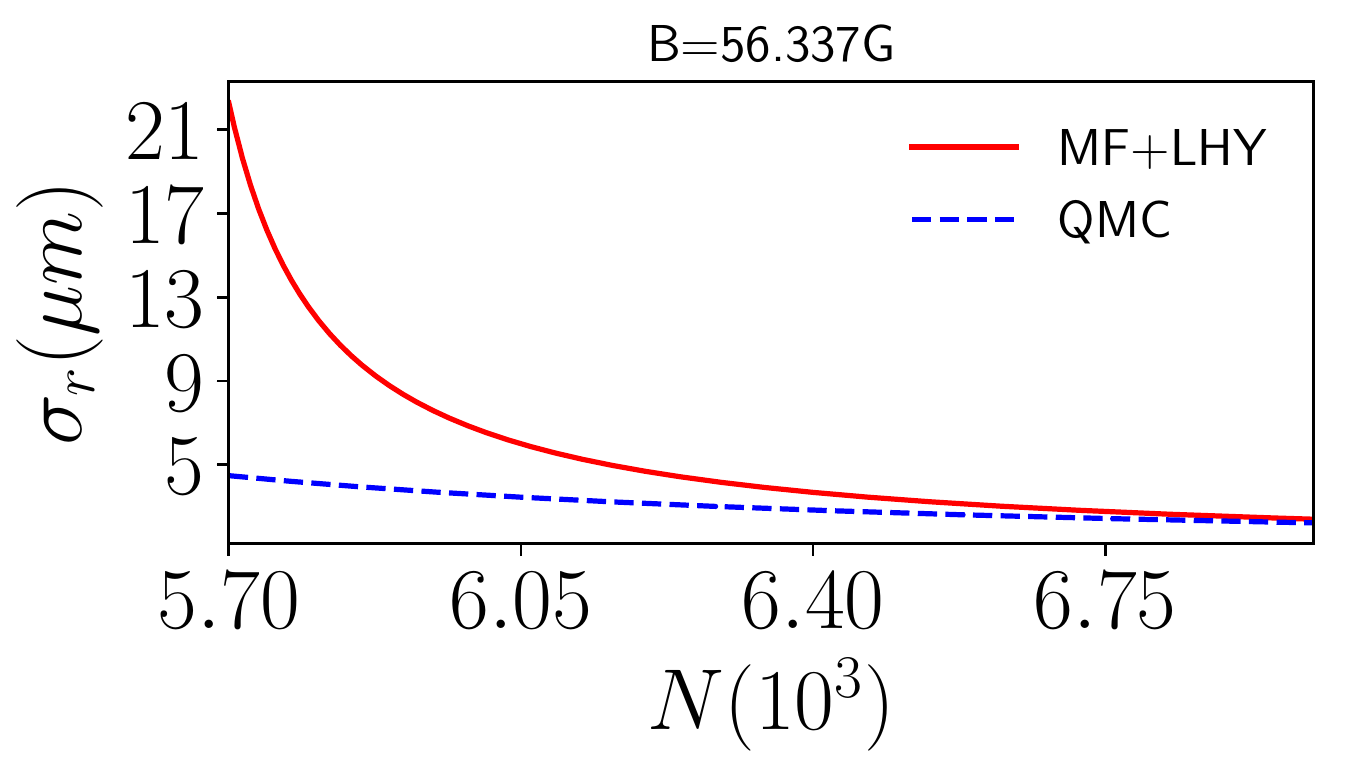}
		\caption{Dependence of the radial size $\sigma_r$ on the 
number of particles. The size is obtained from the variational ansatz, since 
close 
to the critical atom number the density profile in the radial direction is well 
approximated by a Gaussian. In both functionals, it is assumed that the 
relative 
concentration is optimal $N_2 / N_1 = \sqrt{a_{11} / a_{22}}$. QMC functional includes the correct finite-range $r_{\rm eff}$ through POT. 1 set of potentials, Fig.  \ref{fig:ploteosfinitesizeb56337}.}
		\label{fig:sigmarvsncompareqmcmflhy}
	\end{figure}

	As we can see in Fig.~\ref{fig:xofbfitcombined}, the value for $x$ becomes 1 (optimal value) when the drop composed by 15000 particles is studied at the highest magnetic field. This can be understood if we observe that the critical number for this magnetic field matches approximately this number of atoms (see Fig.\ref{fig:criticalncomparisonwicfo}). When the number of atoms of 
a drop is larger than the critical number (lower $B$ in 
Fig.~\ref{fig:sigmarcompwicfoxofb}) $x$ departs from one. This can be better 
understood if one calculates the drop phase diagram as a function of $x$.
The result is plotted in Fig.~\ref{fig:xnc}. As the number of particles is 
approaching the critical one, the range of possible values of $x$, which 
support 
a drop state, is reducing. This is a supporting fact that drops close to 
the critical atom number observed in the experiment fulfill the condition 
$x=1$. 
On the other hand, there is an increasing range of relative particle concentrations for which a drop can emerge as the number of particles increases.

	Close to the critical atom number, the density profile of a drop can 
change 
drastically depending on the functional. We illustrate this effect in 
Fig.~\ref{fig:sigmarvsncompareqmcmflhy} for a magnetic field $B=56.337$ G. 
In the figure, we show the dependence of the radial size on the number of particles, with the same  harmonic confinement strength as in one of the experiments~\cite{bb_mixture_first_taruell}. We observe a substantial difference between the MF+LHY and QMC functional results, mainly when $N$ 
approaches the critical number $N_c$.

	\section{\label{discussion} Discussion}
Experiment in Ref.~\cite{bb_mixture_first_taruell} showed a significant 
disagreement between the measured data and the MF+LHY perturbative approach. In 
order to determine the possible origin of these discrepancies we have pursued a 
beyond  MF+LHY theory which incorporates explicitly the finite range of the 
interaction. To this end, we have carried out DMC calculations of the bulk 
liquid to estimate accurately its equation of state. We have observed that the 
inclusion in the model potentials of both the s-wave scattering length and the effective range produces a rather good universal equation of state in terms of 
these pair of parameters. Excluding the effective range, significant differences are obtained from these universal results. This relevant result points to the loss of universality in terms of the gas parameter in the study of these dilute liquid drops.

Introducing the DMC equation of state into the new functional, following the 
steps which are standard in other fields, such as DFT in liquid 
Helium~\cite{barranco_review}, we derive a new functional that allows for an 
accurate study of the  most relevant properties of the drops. In particular, 
we observe that the inclusion of finite range effects reduces the critical atom 
number in all the magnetic field range approaching significantly the 
experimental values. On the other hand, our QMC functional is not able to 
explain the clear discrepancy between theory and experiment about the size of 
the drops. We attribute this 
difference to the dramatic effect on the size that small shifts on the value of 
$x$ 
produce. Our analysis provides a reasonable explanation of this feature: above 
the critical atom number the window of stability of the drops increases from 
the single point $x=1$ to a range of values that, in absolute terms, grow with 
the number of particles. With the appropriate choice of $x$, one can obtain agreement with the experiment.

The drops produced in the different setup of Ref.~\cite{bb_mixture_sec} are 
spherical 
since all magnetic confinement is removed. The corresponding critical numbers 
in this case are larger than in the confined 
setup~\cite{bb_mixture_first_taruell} and MF+LHY theory accounts reasonably 
well 
for the observed features. We have applied our formalism also to this case and 
the corrections are not zero but relatively less important than in the case 
analyzed here.

	\acknowledgments
	We acknowledge very fruitful discussions with Leticia Tarruell, C\'esar 
Cabrera, and Julio Sanz.
	This work has been supported by the Ministerio de
Economia, Industria y Competitividad (MINECO, Spain) under grant No. 
FIS2017-84114-C2-1-P. V. C. acknowledges financial support from STSM Grant
COST Action CA16221.


\begin{thebibliography}{99}
	

	
	
	\bibitem{petrov}  D. S. Petrov, Phys. Rev. Lett. \textbf{115}, 155302 
	(2015).

	\bibitem{petrov_astra2D} D. S. Petrov and G. E. Astrakharchik,
	Phys. Rev. Lett. \textbf{117}, 100401 (2016).
	
	
	
	\bibitem{parisi} L. Parisi, G. E. Astrakharchik, and S. Giorgini, Phys. 
Rev. 
Lett. \textbf{122}, 105302 (2019).

	\bibitem{dipolar_experiment_nature} H. Kadau, M. Schmitt, M. Wenzel, C. 
Wink, T. Maier, I.Ferrier-Barbut, and T. Pfau, Nature (London) \textbf{530}, 
194 (2016).

	\bibitem{raul} R. Bombin, J. Boronat, and F. Mazzanti, Phys. Rev. Lett. 
\textbf{119}, 250402 (2017).


	\bibitem{bb_mixture_first_taruell} C.R. Cabrera, L. Tanzi, J. Sanz, B. 
Naylor, P. Thomas, P. Cheiney, and L. Tarruell, Science \textbf{359}, 301 
(2018).
	
	\bibitem{soliton_to_drop}  P. Cheiney, C.R. Cabrera, J. Sanz, B. Naylor, 
L. Tanzi, and L. Tarruell, Phys. Rev. Lett. \textbf{120}, 135301 (2018).
	
	
	\bibitem{bb_mixture_sec} G. Semeghini, G. Ferioli, L. Masi, C. Mazzinghi, L. 
Wolswijk, F. Minardi, M. Modugno, G. Modugno, M. Inguscio, and M. Fattori, Phys. 
Rev. Lett. \textbf{120}, 235301 (2018).
	 
	
	\bibitem{bb_mixture_het} C. D'Errico, A. Burchianti, M. Prevedelli, L. Salasnich, F. Ancilotto, M. Modugno, F. Minardi, C. Fort, arXiv:1908.00761v2, 2019
	
	
		
	
	\bibitem{barranco_review} M. Barranco, R. Guardiola, S. Hern\'andez, R. 
Mayol, and M. Pi,	J. Low Temp. Phys. \textbf{142}, 1 (2006).
	
	\bibitem{krotscheck_drops} S. A. Chin and E. Krotscheck,
Phys. Rev. B \textbf{52}, 10405 (1995).
	
	\bibitem{jorgensen} Nils B. J\o rgensen, Georg M. Bruun, and Jan J. Arlt, 
Phys. Rev. Lett. \textbf{121}, 173403 (2018).
	
	\bibitem{pethick} C. J. Pethick and H. Smith, \textit{Bose-Einstein 
Condensation in Dilute Gases} (Cambridge University Press, 2008, Cambridge)

	\bibitem{giorgini_boronat} S. Giorgini, J. Boronat, and J. Casulleras, 
Phys. Rev. A \textbf{60}, 5129 (1999).
	
		
	\bibitem{staudinger} C. Staudinger, F. Mazzanti, and R. E. Zillich, Phys. 
Rev. A \textbf{98}, 023633 (2018).
	
	\bibitem{symmetric} V. Cikojevi\'c, L. Vranje\v{s} Marki\'c, G. E. 
Astrakharchik, and J. Boronat, Phys. Rev. A \textbf{99}, 023618 (2019).
	
	\bibitem{tononi1}  A. Tononi, Condens. Matter \textbf{4}, 20 (2019).
	
	\bibitem{tononi2} A. Tononi, A. Cappellaro, and L Salasnich, New J. 
Phys. \textbf{20}, 125007 (2018).
 	
 	\bibitem{salsnich_nonuniversal} Salasnich L., Phys. Rev. Lett. 
\textbf{118}, 130402 (2017).

	\bibitem{scattering_theory_of_waves_and_particles_1}  R. G. Newton, Scattering 
	Theory of Waves and Particles (Springer-Verlag, New York, 1982)
	
	\bibitem{scattering_theory_of_waves_and_particles_2} P. Roman, 
Advanced Quantum Theory (Addison-Wesley, Reading, MA, 1965).
	
	\bibitem{flambaum} V. V. Flambaum, G. F. Gribakin, and C. Harabati, Phys.
	Rev. A \textbf{59}, 1998 (1999).
	
	\bibitem{effective_range_taruell} L. Tanzi, C. R. Cabrera, J. Sanz, P. 
Cheiney, M. Tomza, and L. Tarruell, Phys. Rev. A \textbf{98}, 062712 (2018).
	
		
	\bibitem{cikojevic} V. Cikojevi\'c, K. D\v{z}elalija, P. Stipanovi\'c, L. 
Vranje\v{s} Marki\'c, and J. Boronat, Phys. Rev. B \textbf{97}, 140502(R) 
(2018).
	
		
	\bibitem{kh_theorem} P. Hohenberg and W. Kohn, Phys. Rev. \textbf{136}, 
B864 (1964).
	
	\bibitem{dmc} J. Boronat and J. Casulleras, Phys. Rev. B 	\textbf{49}, 8920 (1994).
	
	\bibitem{jastrow} L. Reatto and G. V. Chester, Phys. Rev. \textbf{155}, 88 
(1967).
	
	
	\bibitem{francesco_manuel} Francesco Ancilotto, Manuel Barranco, Montserrat 
Guilleumas, and Mart\'i Pi, Phys. Rev. A \textbf{98}, 053623 (2018).
	

	
	\bibitem{chin_krotchek} Siu A. Chin, S. Janecek, and E. Krotscheck, Chem.
Phys. Lett.  \textbf{470},  342 (2009).
	

	
	
	\bibitem{manuel_marti} M. Pi and M. Barranco, private communication.

		
	\bibitem{sssb_potential} L. M. Jensen, H. M. Nilsen, and G. Watanabe, 
Phys. Rev. A \textbf{74}, 043608 (2006).

		\bibitem{pade} J. Pade, Eur. Phys. J. D \textbf{44}, 345 (2007).

	
	
	


	\bibitem{sclens} S. Roy, M. Landini, A. Trenkwalder, G. Semeghini, G. 
Spagnolli, A. Simoni, M. Fattori, M. Inguscio, and G. Modugno, Phys. Rev. Lett. 
\textbf{111}, 053202 (2013).
	
	

	
	
	
	
	
\end{thebibliography}
\end{document}